\documentclass[aps,eqsecnum,twocolumn,nofootinbib,prd]{revtex4}
\usepackage{amsmath}
\usepackage{amssymb}
\usepackage{amsbsy}
\usepackage{epsfig}
\usepackage{bm}

\begin{document}

\title{\LARGE Search for monopoles using superconducting quantum interference device (SQUID)}
\author{Y. H. Yuan}
\email[E-mail: ]{henry@physics.wisc.edu}
\date{December 23, 2005}

\begin{abstract}
The discovery of magnetic monopoles would be of fundamental 
significance in the research of modern physics. In this paper, we present a short review of the history of magnetic monopole research. 
The theoretical work and experimental technique in the search for 
magnetic monopoles using SQUID (superconducting quantum interference device) are investigated. 
We also discuss the properties of magnetic monopole and propose neutrinos have magnetic charges and a possible experimental test based upon the Faraday induction method.  
\end{abstract}
\maketitle

\section{Introduction}

The existence of the magnetic monopole was one of the open questions of modern physics. In 1931, P. Dirac made the first convincing proposal that the existence of even a single magnetic monopole may explain the quantization of electric charge.  According to the quantization condition of a system of an electron and a magnetic monopole with charge {g} due to Dirac\cite{d1,d2},
\begin{equation}
eg=n\frac{\hbar c}{2}
\end{equation}
where $\hbar=\frac{h}{2\pi}$, $c$ is the speed of light and $n$ is an integer. We should modify this quantization condition by a factor of 3 if free quarks are found. Moreover, the introduction of magnetic monopole would bring a symmetry to the Maxwell's equations of electromagnetism. Such appealing proposal exhilarated a number of experimental investigations since then. Just as J. D. Jackson has described in his famous book \cite{ja} on Classical Electrodynamics:

{\it ``chiefly because of an early, brilliant 
theoretical argument of Dirac, the search 
for monopoles is renewed whenever a new energy region is opened up in 
high energy physics or a new source of matter, such as rocks 
from the moon, becomes available". }

New efforts to search for magnetic monopoles have been motivated considerably since Grand Unified Theory (GUT) of strong and electroweak interactions predicted the magnetic monopoles as well. In 1974, 't Hooft and Polyakov \cite {th, po} pointed out that a unified gauge theory in which electromagnetism is 
embedded in a semisimple gauge group would predict the existence of the 
magnetic monopole as a soliton with spontaneous symmetry breaking. To be more specific, a semisimple non abelian gauge group may break into its subgroups including U(1) which essentially describes magnetic monopole in the framework of GUT. Numerous experimental searches for magnetic monopoles in cosmic radiation and for magnetic monopoles trapped in matter for example at accelerator have been carried out\cite{al,rr,eb,ki,je,mac,her,cdf}. Thanks to the proposal due to Dirac. Although we have not observed magnetic monopoles, the elusive monopole has found its application to a number of different research areas in physics, such as in particle physics, condensed matter physics, string theory, astrophysics and cosmology.

\section{Detections of magnetic monopoles}

Various techniques of detection in the experiments to search for magnetic monopole have been developed since the 1930s. In this paper, we concentrate on the induction technique by exploying the 
Magnetometer SQUID (superconducting quantum interference device). As Tassie showed in his paper \cite{tass} that when moving through a superconducting loop, a magnetic monopole would 
induce a supercurrent in the loop because of the change in magnetic flux through the loop surface. The 
passage of a magnetic monopole through superconducting loop would result in a magnetic flux change of 2$\phi _0$, where $\phi _0=\frac{hc}{2e}=2\times 10^{-7}$ $G\cdot cm^2$ 
is the flux quantum. To be more specific, let us consider a monopole with 
charge g passing along the axis of a superconducting loop. Due to the Maxwell's equation involving the magnetic monopole current $\vec J_m$, 

\begin{equation}
\frac{1}{c}\frac{\partial {\vec B}}{\partial t}+\nabla\times \vec{E}=-
\frac{4\pi }{c} {\vec {J_m}}.
\end{equation}

By integration and Stokes theorem, we obtain

\begin{equation}
\frac{1}{c}\frac{\,d}{\,dt}{\int \vec{B}\cdot d \vec {S}}
 =-\oint \vec{E}\cdot d \vec {l}-\frac{4\pi}{c}\int \vec {J_m}\cdot d 
\vec{S}
\end{equation}
where path $l$ is the boundary of area $S$. According to the theory of superconductivity\cite{lo}, the 
fluxoid $\phi_c$ is  

\begin{equation}
\phi_c=\int \vec{B}\cdot d \vec {S}+c\oint \Lambda\vec{j}\cdot d \vec 
{l}
\end{equation}
where constant parameter $\Lambda $ is related to the penetration depth $\lambda_L$ 
which is the characteristic length of superconductor.

\begin{eqnarray}
\lambda_L &=& \sqrt \frac{mc^2}{4\pi n' e_s^2} \nonumber \\
&=&\sqrt \frac{\Lambda c^2}{4\pi }
\end{eqnarray}
where $n'$ is the number of the carriers of the supercurrent per unit 
volume, $m$ and $e_s $ are the mass and charge of the carriers of the  
supercurrent respectively. According to the BCS theory, the carriers of 
the supercurrent are Cooper pairs which are paired electrons. So we 
obtain 

\begin{eqnarray}
m&=&2m_e \nonumber \\
e_s&=&2e
\end{eqnarray}
where $m_e$ and $e$ are the mass and electric charge of an electron.
The value of the penetration depth depends on the temperature. 

Following Tassie's assumpation that the initial and final conditions are stationary, therefore when a monopole passing through the superconducting loop the change in the fluxoid is

\begin{equation}
\Delta \phi_c =-4\pi g
\end{equation}

In addition the change in the magnetic flux through the superconducting loop is approximately the same as the change in the fluxoid especially in case that the superconducting loop is sufficiently thick, so we have

\begin{eqnarray}
\Delta\phi &\approx & \Delta \phi_c  \nonumber \\
&=& -4\pi g
\end{eqnarray}

where $\Delta \phi $ denotes the change in the magnetic flux. In experiments, the 
Magnetometer SQUID is needed to monitor the small induced supercurrent in the superconducting loop due to 
a monopole. Shielding from magnetic fields by using superconductors is of great 
importance for the operation of SQUID because the motion of the 
magnetic flux quanta trapped in the SQUID sensor may produce signals mimicking the magnetic monopole 
events. As a matter of fact, in experiments, both SQUID and detector loops should be placed in the 
space bounded by superconducting shields. 

Searches for magnetic monopole should based upon the properties of magnetic monopole. Therefore we summarize the properties of magnetic monopole next. It is an electrically neutral particle unlike dyon, due to Schwinger\cite{sc,sh}, which is both electrically and magnetically charged. It differs from light quanta in not travelling with the speed of light. In addition, magnetic charge density is a pseudoscalar. When looking at a magnetic monopole from both the right-handed coordinate system and the left-handed coordinate system, we find the signs of a magnetic charge are opposite in the two coordinate systems. Therefore the space inversion of an interaction involving a magnetic monopole would be violated. As J. D. Jackson pointed out in his famous book \cite{ja} on Classical Electrodynamics:
``... it is a necessary consequence of the existence of a particle with both 
electric and magnetic charges that space inversion and time reversal are no longer 
valid symmetries of the laws of physics. It is a fact, of course, that these 
symmetry principles are not exactly valid in the realm of elementary 
particle physics, but present evidence is that their violation is 
extremely small and associated somehow with the weak interaction." This behavior of the magnetic monopole is quite similar to the behavior of neutrino\footnote{In this paper, neutrino means electron neutrino only except when specified.}. As a matter of fact, parity violations\cite{ly,wu,ga} always take place in weak interactions whenever there are neutrinos involved. We suggest that the elusive neutrino has magnetic charge. The flavor change of neutrino is the direct consequence of this proposal\cite{yy} and we may also explain the long-standing solar neutrino puzzle easily. Before completing this section, we give a possible experimental test based upon the Faraday\cite{fa} induction method. Place a radioactive source at the center of an enclosed superconducting sphere rather than superconducting loops. Whenever $\beta$-decay of the radioactive source happens, an anti-neutrino is released which would induce the supercurrents on the superconducting sphere due 
to the proposal. A SQUID or a scanning system may be exploited to monitor the 
supercurrents. To eliminate any unwanted influence 
of electrically charged particles, an absorbent layer would be introduced between the 
radioactive source and the enclosed superconducting sphere. Even though, the sensitive devices in the experiment are vulnerable to spurious signals\cite{kl}, it is still an ideal way to detect the monopole since the method is independence of the particle's mass and velocity. 
The detectors should be placed inside a magnetic shield made 
up of lead or mumetal to protect the detectors from external magnetic 
fields. 

\section{Summary and conclusion}

The search for the magnetic monopole would be of fundamental 
significance in modern physics. A great deal of efforts 
have been made to detect magnetic monopole since the prediction of existence of the 
magnetic monopole by Dirac. We have presented a short review of history of magnetic monopoles. 
The theoretical work and experimental technique in the search for 
magnetic monopoles using SQUID are investigated in the present paper. The change in fluxoid 
as well as in magnetic flux have been discussed when a magnetic monopole moves 
through the superconducting loops. We have also studied the properties of magnetic monopole and proposed a possible experimental test based upon the Faraday induction 
method. This year is the unprecedented World Year of Physics which marks the hundredth 
anniversary of the pioneering contributions of Albert Einstein. 
We dedicate this paper to Albert Einstein.

\end{document}